\documentclass{article}
\usepackage{spconf,amsmath,graphicx}
\usepackage{amsmath,amssymb}

\usepackage{enumitem}
\usepackage{float}
\setlist{nosep, leftmargin=14pt}
\usepackage{url}

\usepackage{mwe} 

\title{ACTIVE PROMPT TUNING ENABLES GPT-4o TO DO EFFICIENT CLASSIFICATION OF MICROSCOPY IMAGES}
\name{
\begin{minipage}[t]{\textwidth}
    Abhiram Kandiyana\textsuperscript{1} \quad
    Peter R. Mouton\textsuperscript{2} \quad
    Yaroslav Kolinko\textsuperscript{3} \quad
    Lawrence O. Hall\textsuperscript{1} \quad
    Dmitry Goldgof\textsuperscript{1}
  \end{minipage}
}

\address{\textsuperscript{1} University of South Florida, Tampa, FL;
\textsuperscript{2} SRC Biosciences, Tampa, FL \\
\textsuperscript{3} Charles University, Prague, Czech Republic}

\begin{document}

\maketitle
\begin{abstract}
Traditional deep learning-based methods for classifying cellular features in microscopy images require time- and labor-intensive processes for training models. Among the current limitations are major time commitments from domain experts for accurate ground truth preparation; and the need for a large amount of input image data. We previously proposed a solution that overcomes these challenges using OpenAI's GPT-4(V) model on a pilot dataset (Iba-1 immuno-stained tissue sections from 11 mouse brains). Results on the pilot dataset were equivalent in accuracy and with a substantial improvement in throughput efficiency compared to the baseline using a traditional Convolutional Neural Net (CNN)-based approach.

The present study builds upon this framework using a second unique and substantially larger dataset of microscopy images. Our current approach uses a newer and faster model, GPT-4o, along with improved prompts. It was evaluated on a microscopy image dataset captured at low (10x) magnification from cresyl-violet-stained sections through the cerebellum of a total of 18 mouse brains (9 Lurcher mice, 9 wild-type controls). We used our approach to classify these images either as a control group or Lurcher mutant. Using 6 mice in the prompt set the results were correct classification for 11 out of the 12 mice (92\%)  with  96\% higher efficiency, reduced image requirements, and lower demands on time and effort of domain experts compared to the baseline method (snapshot ensemble of CNN models). These results confirm that our approach is effective across multiple datasets from different brain regions and magnifications, with minimal overhead.
\end{abstract}
\section{Introduction}
Accurate assessments of cellular damage are essential for evaluating brain aging, neurotoxicology, and the efficacy of potential treatments for neurological conditions such as Parkinson’s, Huntington's, and Alzheimer’s diseases. Previous work shows deep learning architectures can be trained to automatically quantify cell loss in defined regions of interest (ROIs) using disector-based multiple-input multiple-output (MIMO)
\begin{figure*}[t]
\label{fig:test-workflow}
\centering
\includegraphics[width=1\linewidth]{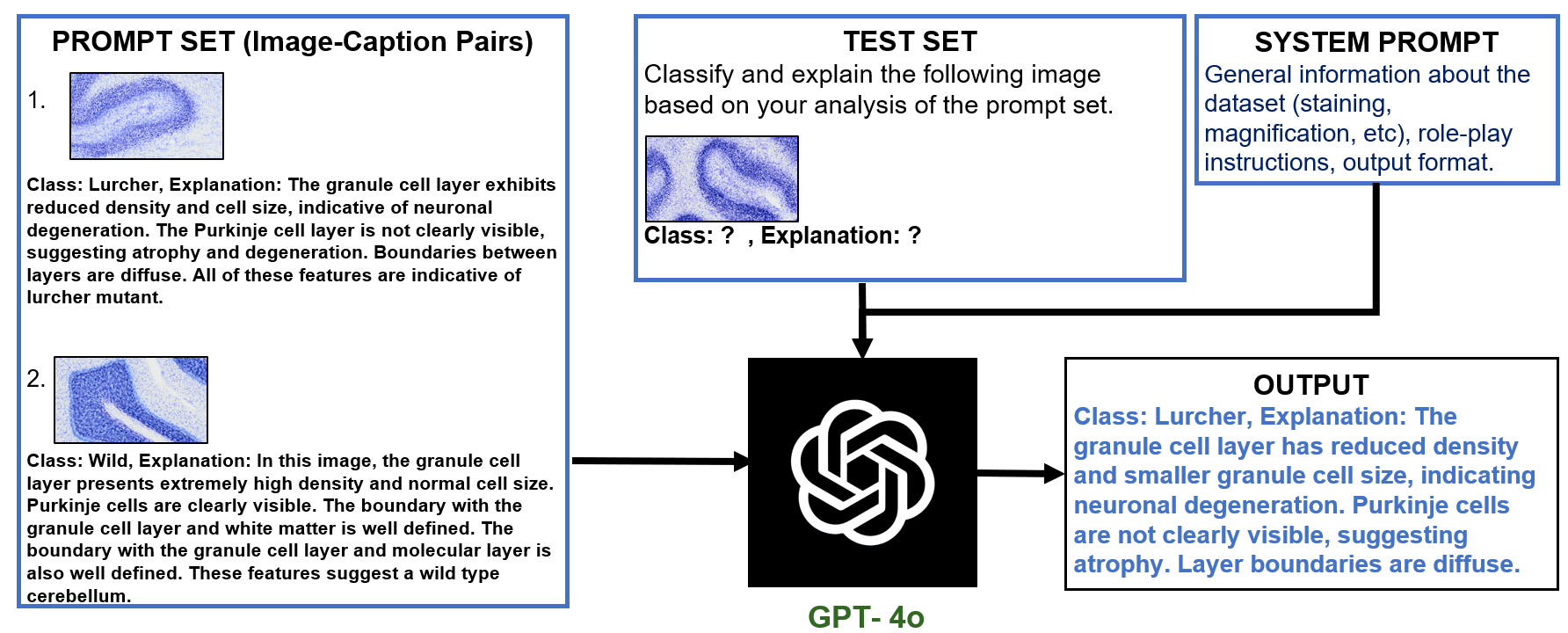}
\caption{\ninept Illustration of the workflow for image classification with GPT-4o. 
}
\end{figure*}
frameworks \cite{palakmimounet,morera2023mimo}. This approach shows comparable accuracy compared to manual stereology counts by humans ($>$90\%) with improved throughput efficiency and less supervised time from experts. Further benefits of these frameworks include enhanced reproducibility (100\% test-retest reliability) due to lower subjectivity and less variation from human factors such as training, experience, fatigue, etc.

A subsequent study \cite{morera20x} demonstrated that snapshot ensembles of CNNs \cite{CNNintro} trained on low-magnification (10-20x) images can make accurate predictions of microglial proliferation at the global level. This CNN-based approach showed a classification accuracy of 10 out of 11 cases correct (91\%). 
A limitation of deep learning-based approaches for automatic cell counts is the requirement for quantitative ground truth, i.e., stereology-based counts, for training separate models for each cell type. Collection of this ground truth data cannot be crowd-sourced due to requirements for domain experts and integrated hardware-software stereology systems. Second, the requirement for labeling microscopy images (annotation) in specific regions of interest is laborious, time-intensive, and often requires a trained technician under the supervision of an experienced domain expert. Third, these approaches require a pre-trained or custom-built model followed by hyper-parameter optimization, training, and testing which can take months of effort by an AI expert. Finally, this labor-intensive, time-consuming, and computationally expensive process must be repeated for every new dataset (cell type). 

The recent introduction of Vision Language Models (VLMs) has the potential to reduce this bottleneck through in-context learning capabilities \cite{sota-vlms,why-ICL} and simple prompting of a pre-trained model with a few examples at inference time, i.e., without the need for extensive fine-tuning. VLMs provide generative AI with the capability to understand, interpret, and analyze both text and images in concert. A further strength of VLMs is their ability to leverage knowledge from one modality to inform the analysis of another. Prior work with VLMs reports state-of-the-art results for general and domain-specific Visual Question Answering (VQA), image-text, text-image, and image-image retrieval \cite{sota-vlms, biomedclip}. Most VLMs incorporate separate encoders for images and text and then use contrastive learning to capture the association between the text, i.e., a sentence, and an image. Learning of cross-modal representations is achieved by maximizing the similarity between matched image-text pairs while minimizing the similarity between unmatched pairs \cite{clip,biomedclip}.

GPT-4 omni (GPT-4o, for short) is the flagship multi-modal model from OpenAI. Following inputs of prompt sets, i.e., text descriptions of class-specific features in corresponding images, GPT-4o generates output in the form of classifications with text-based explanations for subsequent test images. These explanations can be accepted as is, or corrected and used as the prompt set for a subsequent iteration. As of May 2024, GPT-4o is the state-of-the-art model in vision understanding benchmarks, surpassing other leading models like GPT-4(V), Claude Opus, and Gemini 1.0 Ultra \cite{gpt-4o,gpt-4o-radiology}.

Our previous work used the then state-of-the-art VLM model, GPT-4(Vision), for the classification of  Iba-1 immunostained microglia cells in the hippocampus of tissue sections through brains of mice treated with either a powerful neurotoxin (tri-methyl tin, TMT) or saline \cite{cbms_paper}. A pilot study on a subset of data from the current study compared GPT-4(Vision) and GPT-4o. This experiment showed superior performance and reduced latency for GPT-4o. For this reason, we selected GPT-4o for subsequent experiments.

In this work, we evaluated our Active Prompt Tuning \cite{cbms_paper} on a dataset of low-magnification (10x) images captured from mouse brain cerebellum sections stained with cresyl violet, a relatively low signal: noise general marker for all brain cells \cite{kolinko2016}. This dataset included images from 18 mice, a substantially larger number of mice than in our previous work with microglial cells from mouse hippocampus using the same few-shot prompting approach \cite{cbms_paper}. Specifically, we used images from 9 Lurcher mutants and 9 wild-types (controls) to assess the classification accuracy of the model using few-shot prompting. Notably, the number of images used for ground truth training (prompt sets) represents a small fraction (2\%) of the image dataset required for the baseline CNN method. We also contrast the testing accuracy and time required to prepare the ground truth for the classification of two diverse datasets at different magnifications using our approach based on few-shot prompting compared to a traditional CNN approach \cite{morera20x}.

\section{Methods}
The fundamental approach is a VLM for classifying cellular features in microscopy images with minimum reliance on ground truth images, as detailed in the following section.

 All experiments used few-shot prompting techniques of a VLM for classification, rather than fine-tuning or other forms of training as required for CNN models. Few-shot prompting requires only a few examples (prompt sets) of input data with a corresponding textual description of the expected output at inference time \cite{gpt-3,gpt4-covid,fewshotlearning}. The prompt sets provide the context for adapting the vast knowledge in the training base, e.g., the entire Internet, to the specific task at hand without the need for fine-tuning or further training.

 In our experiments during the pilot study, we explored different prompting strategies, ranging from zero-shot prompting to varying levels of manual intervention in few-shot prompting \cite{cbms_paper}. Empirical observations show that some degree of manual correction of the prompts ensures useful results from the model. This observation aligns with recent research findings that emphasize the high sensitivity of VLMs to input prompts \cite{COT-prompting-factors, apple-paper}.

 Although input prompts are required for only a small subset of the dataset, manually curating these prompts is a time- and labor-intensive process that demands experts from the problem domain (neuroscience) and sometimes linguistics. To address this issue, we proposed a novel ``Active Prompt Tuning" step to automate the prompt set selection step as follows \cite{cbms_paper}.

Active prompt tuning (APT) is an active-learning-based human-in-the-loop approach for selecting the most effective prompt set for a given task \cite{cbms_paper}. In the first step, a random subset of images is selected from the dataset according to basic criteria. For example, here we selected 6 random images from each of 6  (out of 18) random mice selected for prompting. This subset was further divided into the ``active set" and the ``initial prompt set". Under the supervision of a highly experienced domain expert (PRM), a short caption was created for each image in the initial prompt set. These short captions provided the model with clear descriptions of the visual cues in the image, enabling it to effectively classify the images in the test set. In this way, the prompt set directs the model's vast knowledge to the task of classifying particular biomedical images and ensures the model generates the output in the expected format. Each caption consists of two parts: 
\begin{itemize}
    \item The ground truth classification of the image
    \item a brief explanation highlighting key morphological features that support its ground truth classification
\end{itemize}
These ``image-caption pairs" provide the initial prompt set.

Second, the GPT-4o model is prompted to generate captions for images in the active set using the initial prompt set as a few-shot prompting example. The model input includes both the image-caption pairs from the initial prompt set and the images from the active set. By applying in-context learning \cite{ICL-examples} to the active set of images, the model generates outputs consisting of two items for each image: 1) the predicted classification (e.g., Lurcher vs. wild-type); and 2) a brief (1-3 sentence) explanation for that classification decision. We will refer to the method used in the pilot study as APT \cite{cbms_paper} and the current version of our method with improved prompts and the newer and faster GPT-4o model \cite{gpt-4o} as APT-USF throughout the remainder of this paper.

Third, the correctly classified images from the active set are reviewed by a human expert who verifies and, if necessary, corrects and/or refines the captions. Once verified, the image-caption pairs are added to the initial prompt set. This process repeats for multiple rounds, with each round potentially adding more correct and detailed classified samples from the active set to the initial prompt set. The rounds continue until all images in the active set are correctly classified and moved to the initial prompt set. In cases where certain images in the active set are not predicted correctly after several rounds, a threshold (5 rounds) is set to stop the process and prevent it from running indefinitely.

Once APT-USF is complete, the initial prompt set becomes the “effective prompt set”, consisting of 36 images, which is then used to prompt the model for predictions on the yet unseen test set. This approach significantly reduces the manual overhead as only the initial prompt set requires detailed ground truth preparation. For the active set, ground truth preparation is reduced to a verification step, and thus a major improvement in efficiency.

In addition to the prompt set, we developed a “system prompt” to standardize our interactions with the GPT-4 model for all images. This prompt contains both task-specific instructions and general information about the dataset, ensuring consistency in the model’s outputs. Specifically, the prompt provides contextual details about the images, including magnification, staining method, and anatomical features that apply to all samples. The prompt requests the model to "role-play" as an expert in morphological analyses to enhance the precision of its responses. Furthermore, the model is instructed to classify the images based on characteristics visible in the images belonging to each class. Lastly, specific instructions are provided to ensure the model generates output in a consistent, programmatically parsable format. This system prompt not only improves the model’s performance but also ensures the application of consistent criteria to all images in the dataset.  Figure \ref{fig:test-workflow} shows an illustration of the workflow for microscopy image classification using GPT-4o.

The test set is divided into batches to comply with OpenAI request rate limits. Each API request includes the system prompt, the prompt set, and the test batch, in that order. The model output is parsed into a JSON file for further analysis.

\section{Experiments and Results}
\label{sec:results}

\subsection{Dataset}
\label{ssec:dataset}
The dataset for this work consists of 2-D microscopy images of histologically stained 3-D structures in tissue sections through the cerebellum of 18 mice brains (9 Lurcher mutation, 9 wild-type control group). The classification task involves distinguishing Lurcher mutant mice from the wild type. All the images are captured at low magnification (10x) and stained with cresyl violet, a generic stain for all brain cells. Images from a random subset of 6 mice (3 Lurcher and 3 control) were used for prompting the GPT-4o model while the other 12 mice (6 Lurcher mutation, 6 wild-type controls) were used for testing. The test set contained a total of 1471 images.

\begin{table}[h]
  \centering
   \caption{\ninept Classification performance of our approach on test animals. The second column from the right shows the number of images from each animal classified by the model as Lurcher or wild-type, respectively (e.g., 48/2 means 48 images were predicted as Lurcher and 2 as wild-type). The predicted class for each mouse is determined by majority voting where the final class reflects the highest number of predicted images for that mouse.}
   \vspace{0.5 pt}
   \ninept
   \begin{tabular}{|p{1.4cm}|p{1.8cm}|p{2cm}|p{1.5cm}|}
 \hline
  Test Animal ID&Ground Truth Class& \# Predictions (Lurcher/Wild)& Predicted Class\\
 \hline
 5917&Lurcher&48/2&Lurcher\\
 6323&Lurcher&39/6&Lurcher\\
 \textbf{6350}&\textbf{Lurcher}&\textbf{24/50}&\textbf{Wild}\\
 6480&Lurcher&50/0&Lurcher\\
 6481&Lurcher&38/0&Lurcher\\
 6509&Lurcher&61/0&Lurcher\\
 5973&Wild&1/171&Wild\\
 6132&Wild&0/202&Wild\\
 6134&Wild&4/171&Wild\\
 6349&Wild&5/251&Wild\\
 6353&Wild&2/135&Wild\\
 6483&Wild&16/195&Wild\\
 \hline
\end{tabular}
\label{tab:main-results}
\end{table}


\begin{table}[h]
  \centering
   \caption{\ninept Comparison of accuracy and ground-truth annotation time (in minutes) for APT and Baseline.}
   \vspace{0.5 pt}
   \ninept
   \begin{tabular}{|c|c|c|c|}
 \hline
  Method& Accuracy (\%)& Time&Improvement (\%) \\
 \hline
 APT&91&92&86\\
 Baseline&91&660&N/A\\
 \hline
\end{tabular}
\label{tab:comp-1}
\end{table}

\begin{table}[H]
  \centering
   \caption{\ninept Comparison of accuracy and ground-truth annotation time (in minutes) for APT-USF and Baseline. *  The ground-truth annotation time for the baseline was estimated based on our prior experiments with similar datasets from different brain regions.}
   \vspace{0.5 pt}
   \ninept
   \begin{tabular}{|c|c|c|c|}
   \hline
  Method& Accuracy (\%)& Time&Improvement (\%) \\
 \hline
 APT-USF&92&45&96\\
 Baseline& -- &1080*&N/A\\
 \hline
\end{tabular}
\label{tab:comp-2}
\end{table}

\subsection{Baseline}
Before implementing vision-language models with few-shot prompting for microscopic image classification, our team employed a snapshot ensemble based on a CNN architecture inspired by VGG-16 \cite{morera20x}. 
This baseline model serves as a comparison point for evaluating the relative efficiency and ground truth preparation time in this study.

\subsection{Result Analysis}

The classification results of our approach on the 12 test animals are presented in Table\ref{tab:main-results}. A total of 11 of 12 mice were predicted correctly with a significant margin of correct predictions for all 11 mice, resulting in an overall accuracy of 92\%. 

Tables 2 and 3 provide a comparison between the accuracy and ground-truth annotation time for the APT and APT-USF methods against their respective baseline methods. The `Improvement (\%)' column denotes the percentage reduction in time taken by the APT method compared to the Baseline method, calculated as: \( \frac{\text{Baseline Time} - \text{APT Time}}{\text{Baseline Time}} \times 100 \). Both methods, APT and APT-USF, demonstrate high classification accuracy, with APT achieving 91\% accuracy (Table 2) and APT-USF achieving 92\% accuracy (Table 3). APT, introduced in a previous paper, demonstrated an 86\% reduction in annotation time compared to the baseline (Table 2) \cite{cbms_paper}. However, APT-USF, the newer version of APT introduced in this work, achieves a 96\% reduction in annotation time (Table 3), despite being evaluated on a different dataset with more challenging characteristics (lower magnification).

\section{Conclusion and Future Work}
This work demonstrates that the proposed few-shot prompting framework (APT, APT-USF) can efficiently classify images from two diverse microscopy datasets with different characteristics (immunostained microglial cells in the hippocampus and diverse cresyl violet-stained cells in the cerebellum) captured at different magnifications (10x and 20x). These results show our few-shot prompting approach can generalize across diverse datasets while maintaining an average accuracy of 92\% and improving efficiency (annotation time savings) by an average of 91\% compared to the baseline. Furthermore, our approach requires ground truth preparation for only 2\% of the data, significantly reducing the need for the extensive domain-specific fine-tuning required by traditional CNN-based methods \cite{morera20x} or MIMO methods \cite{palakmimounet,morera2023mimo}.
 
Besides classification labels, a further benefit of our approach is the generation of output explanations for each image, a form of highly detailed ground truth. These explanations add value
by training students to associate the specific features in the image with the classification criteria and helping medical experts build trust in AI-based methods. Finally, these explanations provide the basis for preparing ground truth to train other Vision Language Models (VLMs).

In future studies, we will further explore the capabilities of our approach to generalize to different cell types, staining methods, and magnifications. Second, we will assess the repeatability and robustness of our method under variations to input prompts \cite{prompt-set-size}. Another area of research is adapting our method to open-weight VLMs  \cite{biomedclip}. While we chose GPT-4o due to its status as a state-of-the-art VLM \cite{gpt-4o}, its closed-source nature presents certain uncertainties about continuity and cost limitations.

\section{Compliance with ethical standards}
\label{sec:ethics}
The use of archival tissue sections from animals sacrificed for other projects complies with federal regulations regarding the care and use of laboratory animals:
Public Law 99-158, the Health Research Extension Act, and
Public Law 99-198, the Animal Welfare Act which is regulated
by USDA, APHIS, CFR, Title 9, Parts 1, 2, and 3.

\section{Acknowledgments}
\label{sec:acknowledgments}

This work was supported by National Science Foundation Grants (\#1513126, \#1746511, \#1926990) and a Florida High Tech Corridor Grant (ENG203, \#20-10) to SRC Biosciences and the University of South Florida.

\bibliographystyle{IEEEbib}
\bibliography{main}

\begin{thebibliography}{10}
\small

\bibitem{palakmimounet}
P~{Dave}, Y~{Kolinko}, H~{Morera}, K~{Allen}, S~{Alahmari}, D~{Goldgof}, L.O. {Hall}, and P.R. {Mouton},
\newblock ``{MIMO U-Net: efficient cell segmentation and counting in microscopy image sequences},''
\newblock in {\em Society of Photo-Optical Instrumentation Engineers (SPIE) Conference Series}, J.E. {Tomaszewski} and A.D. {Ward}, Eds., 2023.

\bibitem{morera2023mimo}
Hunter Morera, Palak Dave, Saeed Alahmari, Yaroslav Kolinko, Lawrence~O. Hall, Dmitry Goldgof, and Peter~R. Mouton,
\newblock ``Mimo yolo - a multiple input multiple output model for automatic cell counting,''
\newblock in {\em 2023 IEEE 36th International Symposium on Computer-Based Medical Systems (CBMS)}, 2023, pp. 827--831.

\bibitem{morera20x}
Hunter Morera, Palak Dave, Yaroslav Kolinko, Saeed Alahmari, Aidan Anderson, Grant Denham, Chloe Davis, Juan Riano, Dmitry Goldgof, Lawrence~O Hall, et~al.,
\newblock ``A novel deep learning-based method for automatic stereology of microglia cells from low magnification images,''
\newblock {\em Neurotoxicology and Teratology}, vol. 102, pp. 107336, 2024.

\bibitem{CNNintro}
Keiron O'Shea and Ryan Nash,
\newblock ``An introduction to convolutional neural networks,''
\newblock {\em CoRR}, vol. abs/1511.08458, 2015.

\bibitem{sota-vlms}
Jingyi Zhang, Jiaxing Huang, Sheng Jin, and Shijian Lu,
\newblock ``Vision-language models for vision tasks: A survey,''
\newblock {\em IEEE Transactions on Pattern Analysis and Machine Intelligence}, 2024.

\bibitem{why-ICL}
Damai Dai, Yutao Sun, Li~Dong, Yaru Hao, Shuming Ma, Zhifang Sui, and Furu Wei,
\newblock ``Why can gpt learn in-context? language models implicitly perform gradient descent as meta-optimizers,''
\newblock {\em arXiv preprint arXiv:2212.10559}, 2022.

\bibitem{biomedclip}
Sheng Zhang, Yanbo Xu, Naoto Usuyama, Hanwen Xu, Jaspreet Bagga, Robert Tinn, Sam Preston, Rajesh Rao, Mu~Wei, Naveen Valluri, et~al.,
\newblock ``Biomedclip: a multimodal biomedical foundation model pretrained from fifteen million scientific image-text pairs,''
\newblock {\em arXiv preprint arXiv:2303.00915}, 2023.

\bibitem{clip}
Alec Radford, Jong~Wook Kim, Chris Hallacy, Aditya Ramesh, Gabriel Goh, Sandhini Agarwal, Girish Sastry, Amanda Askell, Pamela Mishkin, Jack Clark, et~al.,
\newblock ``Learning transferable visual models from natural language supervision,''
\newblock in {\em International conference on machine learning}. PMLR, 2021, pp. 8748--8763.

\bibitem{gpt-4o}
OpenAI,
\newblock ``Hello gpt-4o,'' \url{https://openai.com/index/hello-gpt-4o/}, 2024,
\newblock Accessed: 2024-10-20.

\bibitem{gpt-4o-radiology}
T.~Oura, H.~Tatekawa, D.~Horiuchi, S.~Matsushita, H.~Takita, N.~Atsukawa, Y.~Mitsuyama, A.~Yoshida, K.~Murai, R.~Tanaka, T.~Shimono, A.~Yamamoto, Y.~Miki, and D.~Ueda,
\newblock ``Diagnostic accuracy of vision-language models on japanese diagnostic radiology, nuclear medicine, and interventional radiology specialty board examinations,''
\newblock {\em Jpn J Radiol}, Jul 2024,
\newblock Epub ahead of print, PMID: 39031270.

\bibitem{cbms_paper}
Abhiram Kandiyana, Peter~R. Mouton, Lawrence~O. Hall, and Dmitry Goldgof,
\newblock ``Active prompting of vision language models for human-in-the-loop classification and explanation of microscopy images,''
\newblock in {\em 2024 IEEE 37th International Symposium on Computer-Based Medical Systems (CBMS)}, 2024, pp. 205--212.

\bibitem{kolinko2016}
Y.~Kolinko, J.~Cendelin, M.~Kralickova, and Z.~Tonar,
\newblock ``Smaller absolute quantities but greater relative densities of microvessels are associated with cerebellar degeneration in lurcher mice,''
\newblock {\em Front Neuroanat}, vol. 10, pp. 35, Apr 2016.

\bibitem{gpt-3}
Tom Brown, Benjamin Mann, Nick Ryder, Melanie Subbiah, Jared~D Kaplan, Prafulla Dhariwal, Arvind Neelakantan, Pranav Shyam, Girish Sastry, Amanda Askell, et~al.,
\newblock ``Language models are few-shot learners,''
\newblock {\em Advances in neural information processing systems}, vol. 33, pp. 1877--1901, 2020.

\bibitem{gpt4-covid}
Ruibo Chen, Tianyi Xiong, Yihan Wu, Guodong Liu, Zhengmian Hu, Lichang Chen, Yanshuo Chen, Chenxi Liu, and Heng Huang,
\newblock ``Gpt-4 vision on medical image classification -- a case study on covid-19 dataset,'' 2023.

\bibitem{fewshotlearning}
Archit Parnami and Minwoo Lee,
\newblock ``Learning from few examples: A summary of approaches to few-shot learning,''
\newblock {\em arXiv preprint arXiv:2203.04291}, 2022.

\bibitem{COT-prompting-factors}
Akshara Prabhakar, Thomas~L. Griffiths, and R.~Thomas McCoy,
\newblock ``Deciphering the factors influencing the efficacy of chain-of-thought: Probability, memorization, and noisy reasoning,'' 2024.

\bibitem{apple-paper}
Iman Mirzadeh, Keivan Alizadeh, Hooman Shahrokhi, Oncel Tuzel, Samy Bengio, and Mehrdad Farajtabar,
\newblock ``Gsm-symbolic: Understanding the limitations of mathematical reasoning in large language models,'' 2024.

\bibitem{ICL-examples}
Yuanhan Zhang, Kaiyang Zhou, and Ziwei Liu,
\newblock ``What makes good examples for visual in-context learning?,''
\newblock {\em Advances in Neural Information Processing Systems}, vol. 36, 2024.

\bibitem{prompt-set-size}
Jiuhai Chen, Lichang Chen, Chen Zhu, and Tianyi Zhou,
\newblock ``How many demonstrations do you need for in-context learning?,''
\newblock in {\em Findings of the Association for Computational Linguistics: EMNLP 2023}, 2023, pp. 11149--11159.

\end{thebibliography}

\end{document}